\documentclass[twocolumn,secnumarabic,amssymb, nobibnotes, aps, prl,superscriptaddress]{revtex4}

\usepackage{graphicx}
\usepackage{epsfig}

\begin{document}

\title{Resonant enhancement of the zero-phonon emission from a color center in a diamond cavity}%

\author{Andrei Faraon}
\email[Correspondence should be sent to A.F.: ]{andrei.faraon@hp.com}
\affiliation{Hewlett Packard Laboratories, 1501 Page Mill Rd., Palo Alto, CA, 94304, USA}

\author{Paul E. Barclay}
\affiliation{Hewlett Packard Laboratories, 1501 Page Mill Rd., Palo Alto, CA, 94304, USA}
\affiliation{Institute for Quantum Information Science and Department of Physics and Astronomy, University of Calgary, Calgary, Alberta, Canada}
\affiliation{National Institute for Nanotechnology, Edmonton, Alberta, Canada}

\author{Charles Santori}
\affiliation{Hewlett Packard Laboratories, 1501 Page Mill Rd., Palo Alto, CA, 94304, USA}

\author{Kai-Mei C. Fu}
\affiliation{Hewlett Packard Laboratories, 1501 Page Mill Rd., Palo Alto, CA, 94304, USA}
\affiliation{Department of Physics, University of Washington, Seattle, WA, 98195, USA}

\author{Raymond G. Beausoleil}
\affiliation{Hewlett Packard Laboratories, 1501 Page Mill Rd., Palo Alto, CA, 94304, USA}

%


\begin{abstract}

We demonstrate coupling of the zero-phonon line of individual nitrogen-vacancy centers and the modes of microring resonators fabricated in single-crystal diamond. A zero-phonon line enhancement exceeding ten-fold is estimated from lifetime measurements at cryogenic temperatures. The devices are fabricated using standard semiconductor techniques and off-the-shelf materials, thus enabling integrated diamond photonics.

\end{abstract}

\maketitle


Integrated quantum photonic technologies are key for future applications in quantum information~\cite{ref:nielsen2000qcq,ref:obrien2009pqt}, ultra-low-power opto-electronics~\cite{ref:2009.mabuchi.pra.cqedm}, and sensing~\cite{ref:balasubramanian2008nim}. As individual quantum bits, nitrogen-vacancy (NV) centers in diamond are among the most attractive solid-state systems identified to date, owing to their long-lived electron and nuclear spin coherence, and capability for individual optical initialization, readout and information storage~\cite{ref:jelezko2004ocs,ref:balasubramanian2009usc,ref:santori2006cpt,ref:2010.Buckley.ssm,ref:gurudevdutt2007qrb}. The major outstanding problem is interconnecting many NVs for large-scale computation. One of the most promising approaches is to couple them to optical resonators, that enhance the zero-phonon line (ZPL) emission, and can be further interconnected in a photonic network~\cite{ref:cabrillo1999ces,ref:childress2005ftq,ref:togan2010qeb}.

Previous efforts to couple NV centers to optical resonators have been hindered by difficulties with single-crystal diamond fabrication~\cite{ref:2007.wang.fsscd,ref:greentree2006ccd,ref:babinec.2010.FIBDiamond} or integration of diamond with other optical materials~\cite{ref:larsson2009com,ref:fu2008cnv,ref:barclay2009cbm}. Coupling to microresonators has been observed using diamond nanoparticles but the spectral and coherence properties of NVs in these structures is not appropriate for quantum information applications ~\cite{ref:WoltersBenson.APL.2010,ref:barclay2009cie,ref:park2006cqd,ref:2010.Englund.NanoLett.NVtoPC}. Enhanced zero-phonon line emission by coupling to surface plasmons~\cite{ref:2009.kolesov.natphys.wpd,ref:2009.Schietinger.pespe} has recently been reported, but in this case both the ZPL and the phonon sidebands are enhanced. In our approach, the resonator is fabricated directly in a single crystal diamond membrane thus enabling selective enhancement of the ZPL while the emission rate into phonon sidebands remains almost unchanged. By using single crystal diamond it is expected that NV centers with excellent spectral properties could be integrated into an optical quantum network.

\begin{figure}[htp]
\centering
\includegraphics[width=3.1in]{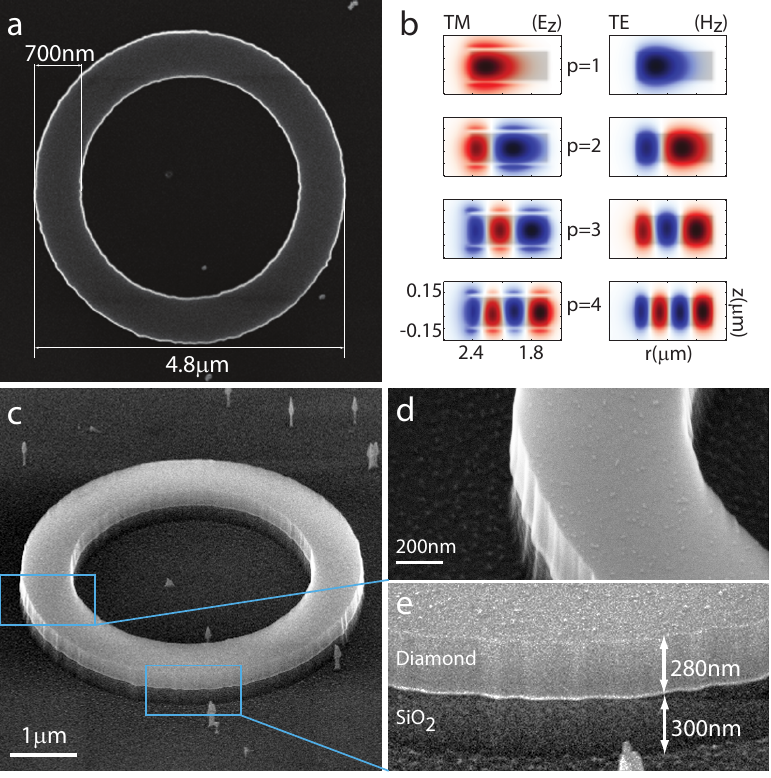}
\caption{{\bf Microring resonator fabricated in single-crystal diamond,} {\bf a}, Scanning electron microscope (SEM) image of the device (top view). {\bf b}, Simulated field profile (azimuthal crosssection) of the TM$_{m=40}$ (E$_z$ field component shown)  and TE$_{m=40}$ (H$_z$ field component shown)
resonances supported by this microring. The $z$ direction is perpendicular to the plane of the ring, $m$ is the azimuthal quantum number and $p$ is the radial quantum number. {\bf c,d,e}, Side view of the microdisk showing the surface roughness due to the fabrication process. }
\label{Fig_device}
\end{figure}

\begin{figure*}[htp]
\centering
\includegraphics[width=6in]{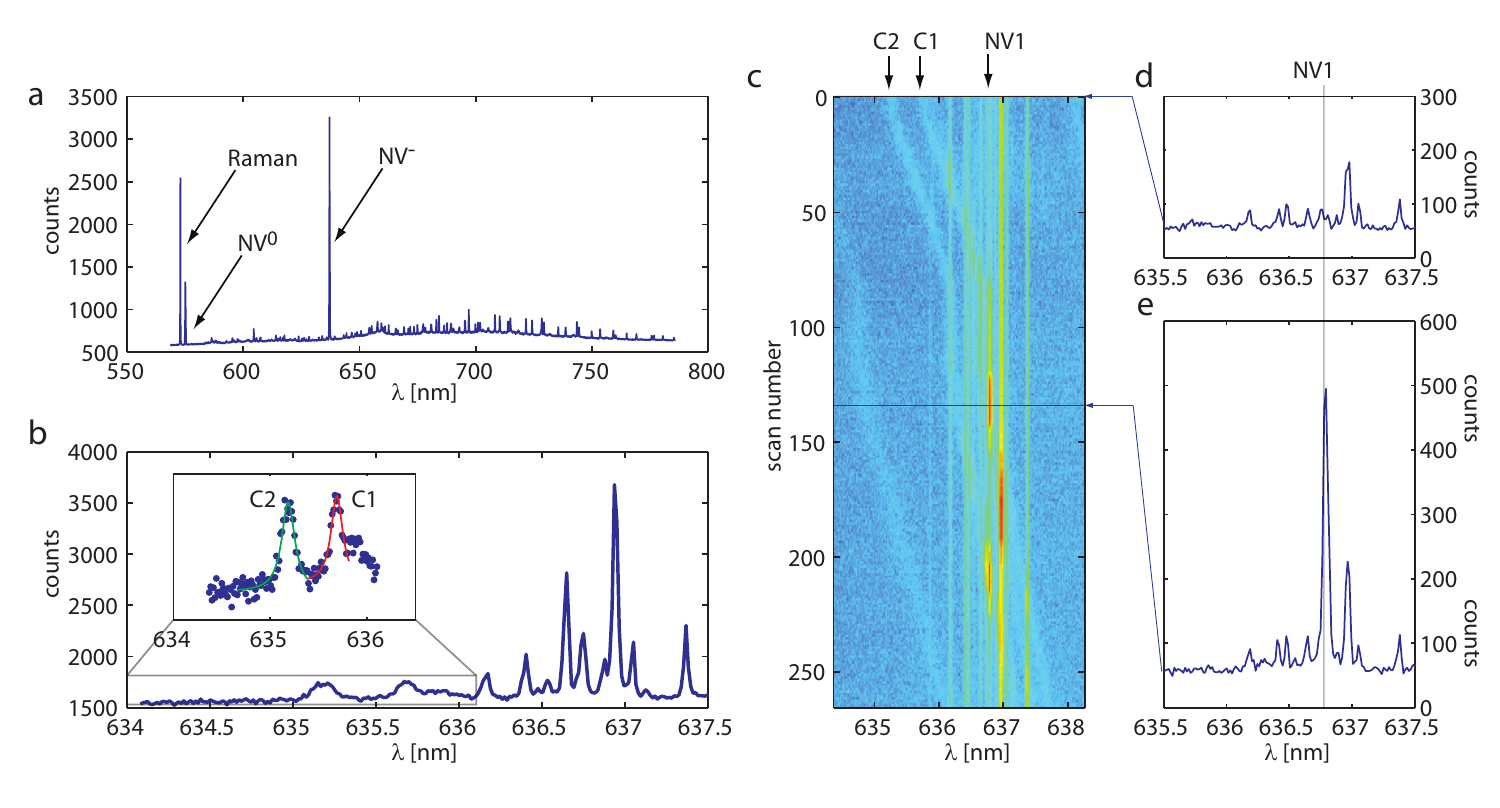}
\caption{ {\bf Spectral characterization showing enhanced photoluminescence due to coupling to the cavity mode.} {\bf a}, Broadband photoluminescence spectrum under continuous-wave excitation. The Raman scattering, and the ZPL from NV$^{0}$ and NV$^{-}$ are marked on the spectrum. Multiple cavity modes can be observed in the phonon sidebands. {\bf b}, Higher resolution spectrum over the NV$^{-}$ ZPL region shows two modes, C1 and C2, blue detuned by only a few nanometers from the ZPL. The quality factors are Q1=4300 and Q2=3800 (see inset). {\bf c}, When tuning C1 and C2 over the ZPL a few NVs exhibit enhanced emission. The tuning rate was reduced when C1 and C2 crossed the ZPL. The transition labeled `NV1' couples to both C1 and C2. The color plot was generated using logarithmic scale.  {\bf d,e}, Spectra of NV1 in the uncoupled ({\bf d}) and coupled case ({\bf e}).}
\label{Fig_spectra}
\end{figure*}

The spontaneous emission rate enhancement of a particular dipole transition $i$ of an emitter coupled to a microresonator relative to the uniform dielectric medium of the resonator is enhanced by the factor $\left( \frac{\tau_{0}}{\tau_{\mathrm{leak}}} \right)_i + F$, where $1/\tau_{0}$ is the emission rate in the uniform dielectric medium, $1/\tau_{leak}$ is the emission rate outside the cavity mode, and
\begin{equation}
F=F_{\mathrm{cav}}\left(\frac{\vec{E}(\vec{r}_i)\cdot\vec{\mu_i}}{\left|\vec{E}_{\mathrm{max}}\right|\left|\vec{\mu_i}\right|}\right)^2 \frac{1}{1+4Q^2(\frac{\lambda_i}{\lambda_{\mathrm{cav}}}-1)^2},
\label{eq:purcell}
\end{equation}
\noindent where $\vec{\mu_i}$ is the dipole moment, $\vec{E}(\vec{r}_i)$ is the local electric field at the emitter location $\vec{r}_i$, $\lambda_{\mathrm{cav}}$ is the cavity wavelength, $\lambda_{\mathrm{i}}$ is the emitter wavelength, and $\left|\vec{E}_{\mathrm{max}}\right|$ is the maximum value of the electric field in the resonator. For the case where the dipole is resonant with the cavity and also ideally positioned and oriented with respect to the local electric field, $F=F_{\mathrm{cav}}$ where,

\begin{equation}
F_{\mathrm{cav}}=\frac{3}{4 \pi ^2} \left(\frac{\lambda_{\mathrm{cav}}}{n}\right)^3\frac{Q}{V_{\mathrm{mode}}},
\label{eq:purcell_cav}
\end{equation}

\noindent and $V_{\mathrm{mode}}=\left( \int_{V}\epsilon(\vec{r}) \left| \vec{E} (\vec{r}) \right| ^2 d^3 \vec{r} \right) / \mathrm{max} \left(\epsilon(\vec{r}) \left| \vec{E} (\vec{r}) \right| ^2 \right) $ is the optical mode volume of the resonator.

The optical cavity used in the experiment consists of a diamond microring on a silicon dioxide pedestal (Fig.~\ref{Fig_device}). The microring, $4.8 \, \mu \mathrm{m}$ in diameter and $700\,\mathrm{nm}$ wide, was etched in a $280\,\mathrm{nm}$ thick diamond membrane that was obtained by thinning a $5\,\mu \mathrm{m}$ single-crystal membrane (Element 6) using reactive ion etching (RIE) in an oxygen plasma. During the etching process the membrane was mounted on a $2\,\mu\mathrm{m}$ thick SiO$_2$ substrate thermally grown on a Si wafer. After the membrane preparation, a silicon nitride ($500\,\mathrm{nm}$) layer was deposited on top and the ring was patterned in this layer using electron-beam lithography and RIE. Two more etching steps were used to transfer the pattern from silicon nitride to diamond and then remove the excess silicon nitride. During the last RIE step, the oxide substrate was also etched by $300\,\mathrm{nm}$. This is a multimode resonator that supports four radial TM modes and four TE modes with electromagnetic field profile shown in Fig.~\ref{Fig_device} {\bf b}. The simulated quality factor can exceed $Q=10^6$, but for the device used in this experiment the quality factor is limited to 5000 by imperfections in the fabrication process, mainly the surface roughness shown in Fig.~\ref{Fig_device} {\bf d,e}. These values of Q are much higher than those observed in microresonators fabricated in nanocrystalline diamond ~\cite{ref:wang2007fct}. The optical mode volume ranges from $V_{\mathrm{mode}} \approx 17 \left(\frac{\lambda}{n} \right) ^3$ to $V_{\mathrm{mode}} \approx 32 \left( \frac{\lambda}{n} \right) ^3$ depending on the particular standing wave mode, where $\lambda$ is the resonance wavelength in air and $n=2.4$ is the index of refraction of diamond. Our simulations indicate the smallest mode volumes for fundamental TE/TM radial modes (p=1) with azimuthal quantum number $m=46$ at $\lambda_{cav}=637\,\mathrm{nm}$.

\begin{figure*}[htp]
\centering
\includegraphics[width=6in]{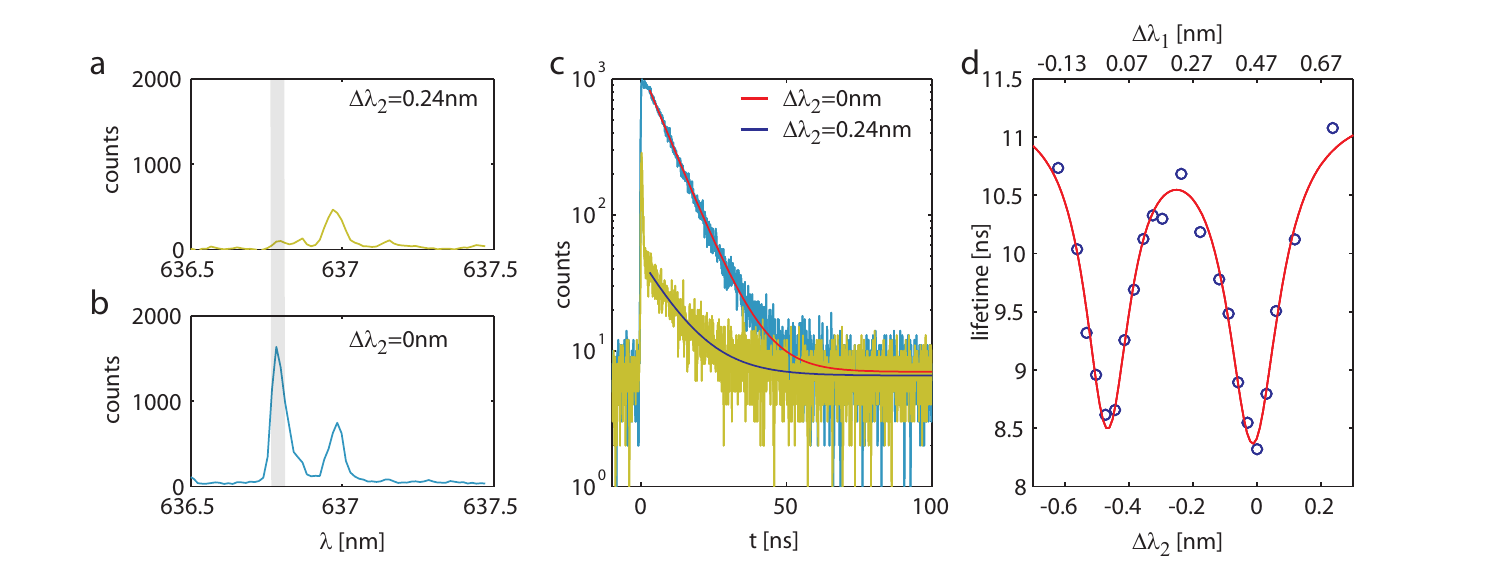}
\caption{{\bf Lifetime measurements for NV1.} {\bf a,b}, Photoluminescence spectrum of NV1 under pulsed excitation when coupled to C2 ($\Delta \lambda _{2}=0\,\mathrm{nm}$) and the uncoupled case ($\Delta \lambda _{2}=0.24\,\mathrm{nm}$). The gray bar marks the spectral filter window used to collect only photons emitted in the NV1 ZPL. {\bf c} Lifetime measurement for the uncoupled ($\tau _{0}=11.1\,\mathrm{ns}$, $\Delta \lambda _{2}=0.24\,\mathrm{nm}$) and the coupled case ($\tau _{c2}=8.3\,\mathrm{ns}$, $\Delta \lambda _{2}=0$). {\bf d}, Change in lifetime as C1 and C2 are tuned over the NV1 ZPL. The data are well fit (red line) assuming coupling to two cavity modes  with quality factors Q1=4300 and Q2=3800 as measured from the spectra, and considering that the emission rate into each cavity mode follows a Lorentzian dependence with respect to detuning.}
\label{Fig_lifetime}
\end{figure*}

During the experiment, the sample was cooled below 10K in a continuous-flow liquid-helium cryostat. A confocal microscope setup was used for optical excitation and collection. The NV centers studied were present in the original membrane, accidentally incorporated during diamond growth. The photoluminescence spectrum while pumping the microdisk with a continuous-wave green laser($532\,\mathrm{nm}$, $2\,\mathrm{mW}$) is shown in Fig.~\ref{Fig_spectra}{\bf a}. Besides the diamond Raman line, the zero-phonon lines and the phonon sidebands from both the neutral (NV$^{0}$) and negatively charged (NV$^{-}$) centers are visible in the spectrum, and multiple cavity modes can be observed on the top of the phonon sidebands. A higher-resolution spectrum (Fig.~\ref{Fig_spectra}{\bf b}) shows that the laser spot with an area of $1 \,\mu\mathrm{m}^{2}$ excites about ten distinct NV$^{-}$ ZPL lines distributed over $\sim 1\,\mathrm{nm}$ because of strain in the sample.  Two resonant modes C1 and C2 with quality factors Q1=4300 and Q2=3800 are spectrally detuned by only a few nanometers from the ZPL at 637nm (Fig.~\ref{Fig_spectra}{\bf b}). Considering a quality factor Q=4000, the maximum expected enhancement factor for an optical dipole ideally aligned and positioned with respect to the local electric field ranges from $1+F \approx 10$ to $1+F \approx 19$. Here, and in the rest of the paper, we assume $\frac{\tau_0}{\tau_{\mathrm{leak}}}=1$, based on our experimental finding that the total decay rate into modes outside of the cavity ($1/\tau_\mathrm{leak}$) is approximately equal to the total decay rate in bulk diamond $1/\tau_0 \approx (12\,\mathrm{ns})^{-1}$.

The modes of the microdisk can be red-shifted by injecting xenon gas into the cryostat. The xenon condenses on the resonator  thus increasing the effective microring size, resulting in a resonance shift. When modes C1 and C2 are tuned over the ZPL a few NV lines show strong intensity enhancement, a signature of coupling to the cavity (Fig.~\ref{Fig_spectra} {\bf c,d,e}). One of the NV lines, marked as NV1, couples to both C1 and C2. These two modes tune at the same rate and have roughly the same quality factor, suggesting that they represent standing waves formed from counter-propagating modes of the same radial and azimuthal order, coupled through surface roughness scattering. The position of the scattering centers also defines the spatial position of the standing waves. The enhancement of the collected photoluminescence depends not only on the emission rate of photons into the cavity mode, but also on the collection efficiency for light emitted into the cavity, and then scattered out.  To obtain a more quantitative estimate of the rate of photon emission into the cavity, we performed photoluminescence lifetime measurements as described below.

The lifetime measurements were performed using green pulsed excitation with a repetition rate of 4.75 MHz using a filtered supercontinuum source ($520\,\mathrm{nm}$ center wavelength, $28\,\mathrm{nm}$ bandwidth, $200\,\mu\mathrm{W}$). The pulsed excitation spectra of NV1, both in the uncoupled case and when coupled to C2, are shown in Fig.~\ref{Fig_lifetime}{\bf a,b}. As expected, under pulsed excitation more photons are collected from NV1 when coupled to the cavity. A monochromator was used to spectrally filter ($0.03\,\mathrm{nm} \approx 20\,\mathrm{GHz}$ bandwidth) only the NV1 ZPL, and the filtered signal was sent to a photon counter. The temporal profile of the photoluminescence emission was fitted by a single exponential function (Fig.~\ref{Fig_lifetime}{\bf c}), indicating a change in the NV1 lifetime from $\tau _{0}=11.1\,\mathrm{ns}$ in the uncoupled case to $\tau _{c2}=8.3\,\mathrm{ns}$ when resonant with cavity mode C2. We note that the first $3\,\mathrm{ns}$ following the excitation pulse were skipped in the fits shown in Fig.~\ref{Fig_lifetime}{\bf c} to minimize the influence of a fast luminescence coming from some contamination on the sample acquired during the sample processing. If the total photon emission rate for uncoupled NVs is $1/\tau _{0}=1/\tau _{\mathrm{ZPL}}+1/\tau _{\mathrm{PS}}$, in the coupled case the rate becomes $1/\tau _{c2}=(1+F)/\tau _{\mathrm{ZPL}}+1/\tau _{\mathrm{PS}}$ where $F$, $1/\tau _{\mathrm{ZPL}}$ and $1/\tau _{\mathrm{PS}}$ are the Purcell factor and the spontaneous emission rate into the ZPL and phonon sidebands, respectively. The Purcell factor is then given by $F=(\tau _{0}/\tau _{c2}-1)/\xi_{\mathrm{ZPL}}$, where $\xi_{\mathrm{ZPL}}$ is the branching ratio into the ZPL ($\xi_{\mathrm{ZPL}}=\tau _{0}/\tau _{\mathrm{ZPL}}$). Our own measurements on other diamond samples indicate a branching ratio of $\xi_{\mathrm{ZPL}}\sim 0.03$, while other sources report values ranging from $\xi_{\mathrm{ZPL}}\sim 0.024$~\cite{ref:davies1974vsd} to $\xi_{\mathrm{ZPL}}\sim 0.05$~\cite{ref:2009.njp.siyushev.ltoc}. A branching ratio of $\xi_{\mathrm{ZPL}}\sim 0.03$, gives a total enhancement factor $ 1+F\approx 12$. This estimate neglects any possible non-radiative decay channels that may also be present.  If such non-radiative decay occurs, then the true ZPL enhancement factor could be even higher than the one estimated here. The lifetime of NV1 was also measured at different detunings from C1 and C2 as shown in Fig.~\ref{Fig_lifetime}{\bf d}. Decreased lifetimes are observed when both C1 and C2 cross over the NV1 resonance. The fit in Fig.~\ref{Fig_lifetime}{\bf d} shows that the change in lifetime follows the expected tuning dependence (Eq.~\ref{eq:purcell}) for a dipole coupled to two cavity modes with quality factors Q1=4300 and Q2=3800 (as measured from the photoluminescence spectra).

\begin{figure}[htp]
\centering
\includegraphics[width=3in]{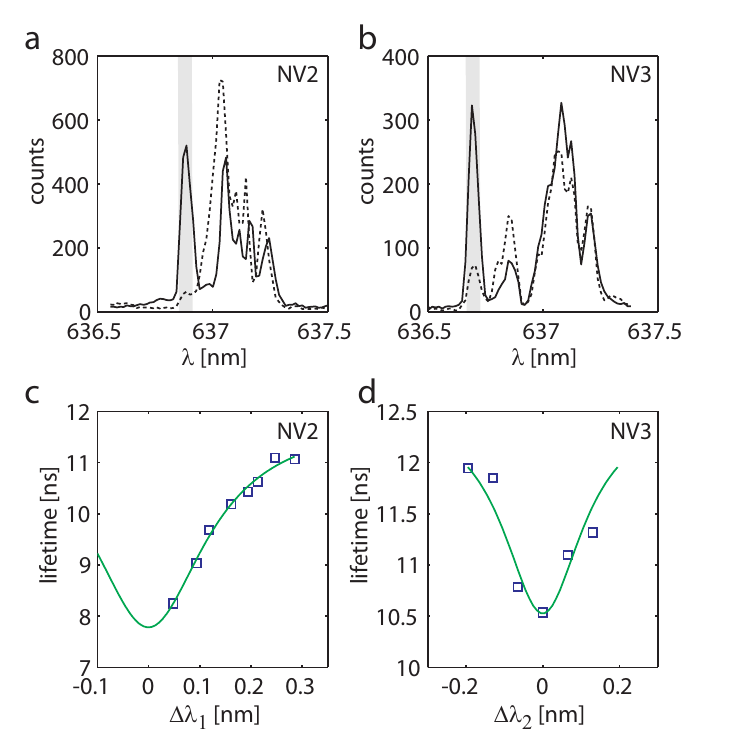}
\caption{{\bf Lifetime measurements for NV2 and NV3.} {\bf a,b}, Photoluminescence spectra for NV2 and NV3 under pulsed excitation both for cavity on resonance with the NV line (continuous), and the cavity off resonant with the NV line (dashed). The gray area shows the spectral filter for the NV ZPL. {\bf c,d} Lifetime as a function of detuning from the cavity mode for NV2 and NV3. }
\label{Fig_lifetime_NV23}
\end{figure}

Other NV lines on this sample showed coupling to cavity modes, as shown in Fig.~\ref{Fig_lifetime_NV23}, for the same microring device. While NV2 was coupled to C1 and showed a similar lifetime reduction as NV1, NV3 was coupled to C2 and on resonance the lifetime was reduced by a relatively smaller amount ($10.4\,\mathrm{ns}$ minimum lifetime). We note that the pulsed photoluminescence spectra in Fig.~\ref{Fig_lifetime_NV23}{\bf a,b} show a higher concentration of NV$^{-}$ than in Fig.~\ref{Fig_lifetime}{\bf a,b}. This difference occurred because the measurements in Fig. 4 (Fig. 3) were performed before (after) the sample was imaged with a scanning electron microscope that caused some conversion from NV$^{-}$ to NV$^{0}$.

The measured spontaneous emission rate enhancement of 12 is smaller than the maximum expected from simulations, because of the non-ideal placement and orientation of the NV with respect to the standing wave. With an enhancement factor of 12, the ZPL branching ratio changes from 3/100 to 36/133 ($\approx 25 \%$). With the same resonator design the branching ratio could exceed 98\% for $Q>5 \times 10^{5}$.  This improvement in Q should be possible with improved fabrication processing. Branching ratios exceeding 99.5\%, as needed for quantum computing protocols, are possible using photonic crystals with $V_{\mathrm{mode}} \approx 2 \left(\frac{\lambda}{n} \right) ^3$ and $Q>2 \times 10^{5}$~\cite{ref:2009.opex.Tomljenovic-Hanic.fduhQ}.

Coupling single NV centers to single-crystal diamond resonators is a first critical step towards large-scale integrated diamond quantum optical networks. In the near term, improvements in fabrication processes should yield structures with higher $Q$, and allow integration of both low-loss waveguides and more complex optical elements such as switches and directional couplers. Studies of the optical manipulation and coherence of single NV spins in cavities will require centers with high-quality spectral properties located at optimal positions with respect to cavity modes. Therefore, in parallel it will be important to develop methods to place NV centers at predetermined positions on a diamond chip~\cite{2010.toyli.nanolett.csn} without introducing additional point defects, and to release the sample strain to reduce inhomogeneous broadening. Even relatively simple chip-scale quantum networks could enable quantum simulators that would significantly outperform classical computers in applications such as quantum chemistry~\cite{ref:lanyon2010tqc}. These devices operating at the fundamental limit of light-matter interaction could also enable future optical signal processing devices and sensors operating at ultra-low power levels, as well as provide a platform for fundamental studies of many-particle entanglement that is a prerequisite of nontrivial quantum factoring and searching~\cite{ref:nielsen2000qcq}.

 This material is based upon work supported by the Defense Advanced Research Projects Agency under Award No. HR0011-09-1-0006 and The Regents of the University of California.


\end{document}